\documentclass[preprint,prd,aps,showpacs]{revtex4}
\usepackage{epsfig}

\begin{document}

\title{Dual parameterization of generalized parton distributions and 
description of DVCS data}

\author{V. Guzey}
\affiliation{Institut f{\"u}r Theoretische Physik II, Ruhr-Universit{\"a}t
Bochum, D-44780 Bochum, Germany}
\email[]{vadim.guzey@tp2.ruhr-uni-bochum.de}

\author{M.V. Polyakov}
\affiliation{Institut f{\"u}r Theoretische Physik II, Ruhr-Universit{\"a}t
Bochum, D-44780 Bochum, Germany and \\
Petersburg Nuclear Physics Institute, Gatchina, St. Petersburg
188350, Russia}
\email[]{maxim.polyakov@tp2.ruhr-uni-bochum.de}

\preprint{RUB-TPII-15/2003}

\begin{abstract}

We discuss a new leading-order parameterization of generalized
 parton distributions of the proton, which is based on the idea of
duality. In its minimal version, the parameterization
is defined by the usual quark  singlet parton distributions 
and the form factors of the energy-momentum tensor.
We demonstrate  that our parameterization describes very well the
absolute value,  the $Q^2$-dependence and the $W$-dependence of the 
HERA data on the total DVCS cross section and contains no free parameters
in that kinematics. The parameterization suits especially well 
the low-$x_{Bj}$ region, which allows us to advocate it as a better alternative 
to the frequently used double distribution parameterization of the GPDs.

\end{abstract}

\maketitle

\section{Introduction}
\label{sec:intro}

Generalized parton distributions (GPDs) have become a standard QCD tool for 
analyzing and parameterizing the non-perturbative parton structure of
 hadronic targets, for
reviews see \cite{Ji:1998pc,Radyushkin:2000uy,Goeke:2001tz,Belitsky:2001ns,
Diehl:2003ny,Belitsky:2005qn}.
In general, GPDs are more general and complex objects than structure 
functions and form factors. In addition, experimentally measured observables
do not access the GPDs themselves but only their convolution with the hard 
scattering coefficients. 
 Therefore, the experimental determination
of the GPDs is an extremely difficult task. Hence, when dealing with the GPDs,
one invariably uses models, the known limiting behavior and 
general properties  of the GPDs and
the physical intuition.

The GPDs have been modeled using virtually all known models of the nucleon 
structure: bag models~\cite{Ji:1997gm}, the chiral quark-soliton 
model~\cite{Petrov:1998kf}, light-front models~\cite{Tiburzi:2001ta,
Tiburzi:2002tq}, constituent quark models~\cite{Scopetta:2003et}, 
Bethe-Salpeter approach~\cite{Tiburzi:2004mh}, a NJL model~\cite{Mineo:2005qr}.
In addition, a double distribution model of the 
GPDs~\cite{Radyushkin:1998es,Musatov:1999xp}
 and modeling by perturbative diagrams~\cite{Pobylitsa:2002vw} were suggested.

The factorization theorem for deeply virtual Compton 
scattering (DVCS)~\cite{Collins:1998be} gives a practical possibility to 
measure the GPDs by studying various processes involving the GPDs:
DVCS, exclusive electroproduction of vector mesons,
wide angle Compton scattering \cite{Radyushkin:1998rt,Diehl:1998kh},
exclusive $p \bar{p} \to \gamma \,\gamma$ annihilation~\cite{Freund:2002cq},
the $p \bar{p} \to \gamma \,\pi^0$ process~\cite{Kroll:2005ni},
$\gamma^{\ast}\, \gamma \to \pi \,\pi$ near threshold~\cite{Diehl:1998dk}.
However, in order to accommodate such a potentially large number of data,
parameterizations of the GPDs should be sufficiently flexible and versatile.
In particular, they should allow for the connection of the 
DVCS with the  $p \bar{p} \to \gamma \,\gamma$ process.

The commonly used double distribution parameterization of the 
GPDs~\cite{Radyushkin:1998es,Musatov:1999xp} is one example of the model 
of the GPDs which could be used to connect
different physical 
channels~\cite{Teryaev:2001qm}.
However, the parameterization of the GPDs based on the double distribution has 
a number of problems from the phenomenological point of view. 
First, in order to have the full form of polynomiality, the so-called 
$D$-term~\cite{Polyakov:1999gs} has to added by hand.
Second, the model dramatically overestimates the low-$x$ HERA data on the 
total DVCS cross section because it involves proton parton distributions at
extremely small and unmeasured values of Bjorken $x$~\cite{Freund:2001hd}.
Third, the model does not allow for an intuitive physical motivation and
interpretation, see~\cite{Freund:2002ff} for a discussion of the physics of GPDs.

In this paper, we offer a new model for the GPDs, which was
in a general form introduced in~\cite{Polyakov:2002wz}.
Unlike the models of the GPDs mentioned above, the present model 
has a simple physical interpretation and direct correspondence 
to the mechanical properties of the target~\cite{Polyakov:2002yz}.
The suggested parameterization of the GPDs can be analytically continued
in $t$ to 
the physical region of the $p \bar{p} \to \gamma \,\gamma$ reaction and
also allows for flexible modeling of the $t$-dependence
of the GPDs.

The considered parameterization of the GPDs is called dual because
the GPDs are presented  as  an  infinite series of 
$t$-channel exchanges, which reminds of the ideas of  duality in hadron-hadron 
scattering. 

In this work, we formulate the minimal version of the 
dual parameterization and determine the free parameters of the model. 
Using the resulting dual parameterization of the GPDs, we successfully describe
the HERA data on the DVCS cross 
section \cite{Adloff:2001cn,Chekanov:2003ya,Aktas:2005ty}.
We explain that our parameterization suits the low-$x_{Bj}$ kinematics 
especially well because the quark singlet parton distributions
 are never probed at the unmeasurably low values
of Bjorken $x$ and because the final expression for the DVCS amplitude
is numerically stable. 
Thus, the dual parameterization of the GPDs gives an opportunity to have
a physically intuitive model of the GPDs, which agrees with the DVCS 
experiments and
which can serve as an alternative to the popular double distribution model.

\section{The dual parameterization of GPDs}
\label{sec:dual}

The dual representation of the GPDs was first introduced for the 
pion GPDs in~\cite{Polyakov:1998ze}. The essence of that derivation is 
presented below. The starting point was the decomposition of the two-pion
distribution amplitude ($2\pi$DA) in terms of the eigenfunctions of the ERBL 
evolution equation (Gegenbauer polynomials $C_n^{3/2}$), 
the partial waves of produced pions (Legendge polynomials $P_l$) and
generalized form factors $B_{nl}$.
The moments of the $2\pi$DA, being the matrix elements of certain local
operators, could be related by crossing to the moments of the pion GPDs.
Then, the pion GPDs could be formally reconstructed using the explicit 
form of their moments.

Based on the result of~\cite{Polyakov:1998ze}, the dual representation
for the proton GPDs was suggested in~\cite{Polyakov:2002wz}.
In this paper, we will consider only the singlet ($C$-even) combination of 
the GPDs $H$, which give the dominant contribution to the total DVCS
cross section at high energies and small $t$. We will work in the leading 
order approximation and, hence, we will consider only quark GPDs.

 The dual representation
of the singlet GPD $H^i$ of the quark flavor $i$ is~\cite{Polyakov:2002wz}
\begin{equation}
H^i(x,\xi,t,\mu^2)=\sum_{n=1,\,{\rm odd}}^{\infty} \sum_{l=0,\,{\rm
even}}^{n+1} B_{nl}^i(t,\mu^2)\, \theta\left(1-\frac{x^2}{\xi^2}\right)\,\left(1-\frac{x^2}{\xi^2}\right)\,
C_n^{3/2}\left(\frac{x}{\xi}\right)\,P_l\left(\frac{1}{\xi}\right) \,,
\label{eq:h}
\end{equation}
where $x$, $\xi$ and $t$ are the usual GPD variables.
The series~(\ref{eq:h}) is divergent at fixed $x$ and $\xi$, and, 
hence, it should be understood as a formal series. In particular, 
it is incorrect to evaluate the series term by term. As a result,
the GPD $H^i$ of Eq.~(\ref{eq:h}) has a support over the entire 
$-1 \leq x \leq 1$ region, 
regardless that each term of the series is non-vanishing only for
$-\xi \leq x \leq \xi$. The formal representation~(\ref{eq:h}) can be 
equivalently rewritten as a converging series using the technique 
developed in~\cite{Polyakov:2002wz}.

The derivation of Eq.~(\ref{eq:h}) used the idea of duality of hadronic 
physics, when the scattering amplitude in the $s$-channel is represented 
as an infinite series of the $t$-channel exchanges. 
This explains the name ``the  dual representation'' 
for the representation of Eq.~(\ref{eq:h}).

As a double series, Eq.~(\ref{eq:h}) is  inconvenient for phenomenological
applications. For the evaluation of the LO DVCS amplitude,
it is useful to introduce the functions $Q_k(x,t)$
 whose Mellin moments generate the $B_{nl}^i$ form 
factors~\cite{Polyakov:2002wz}
\begin{equation}
B_{n\,n+1-k}^i(t,\mu^2)=\int^{1}_{0}dx \, x^n \,Q_k^i(x,t,\mu^2) \,,
\label{eq:qk}
\end{equation}
where $k$ is even. 
A remarkable property of the dual representation is that
the $\mu^2$-evolution of the functions $Q_k^i$ is given by the usual
leading order (LO) DGLAP evolution.

Since the $B_{nl}^i$ form factors are related to the 
moments of $H^i$, the $Q_k$ functions are also constrained by these moments.
In particular, from
\begin{equation}
\int_{-1}^{1} dx \, x \, H^i(x,\xi,t)=M_2^i(t)+\frac{4}{5} \xi^2 d^i(t)=\frac{6}{5} \left[B_{12}(t)-\frac{1}{3} \left(B_{12}(t)-2 B_{10}(t)\right) 
\xi^2 \right] \,,
\end{equation}
it follows that
\begin{eqnarray}
&&\int^{1}_{0}dx \, x \,Q_0^i(x,t,\mu^2)=\frac{5}{6}\, M_2^i(t,\mu^2) \,, \nonumber\\
&&\int^{1}_{0}dx \, x \,Q_2^i(x,t,\mu^2)=\frac{5}{12}\, M_2^i(t,\mu^2)+d^i(t,\mu^2)
 \,,
\label{eq:q02}
\end{eqnarray}
where $M_2^i$ at $t=0$ is the proton light-cone momentum fraction carried
by the quarks; $d^i(t)$ is the first moment of the quark $D$-term.

In addition, the $B_{n n+1}$ form factors at the zero momentum transfer are
fixed by the Mellin moments of the quark singlet parton distribution 
functions (PDFs). In particular,
\begin{equation}
\frac{3}{4}\int^{1}_{0}dx \left(q^i(x,\mu^2)+\bar{q}^i(x,\mu^2)\right)=B_{10}(0)=
 \int^{1}_{0}dx \, \,Q_0^i(x,0,\mu^2) \,.
\end{equation}

The $Q_0^i$ functions at $t=0$ are completely fixed in terms of the 
forward proton PDFs~\cite{Polyakov:2002wz}
\begin{equation}
Q_0^i(x,0,\mu^2)=q^i(x,\mu^2)+\bar{q}^i(x,\mu^2)-\frac{x}{2} \int^{1}_{x}
 \frac{dz}{z^2} \left(q^i(z,\mu^2)+\bar{q}^i(z,\mu^2) \right) \,.
\label{eq:q0}
\end{equation}

As suggested in~\cite{Polyakov:2002wz}, keeping only the functions $Q_0^i$
and $Q_2^i$ constitutes the minimal version of the dual parameterization of
the GPDs. The functions $Q_0^i$ and $Q_2^i$ are defined by Eqs.~(\ref{eq:q0})
and (\ref{eq:q02}), where $M_2^i(t)$ and $d^i(t)$ have a clear physical
interpretation since they are the form factors of the energy-momentum tensor 
evaluated between the states representing the given target.
At $t=0$, $M_2^i(0)$ is the fraction of the plus-momentum of the nucleon 
carried by the quarks of flavor $i$; $d^i(0)$ characterizes the shear forces
experienced by the quarks in the target.

Next we discuss the minimal version of the dual representation in detail.
While $Q_0^i$ at $t=0$ is defined by Eq.~(\ref{eq:q0}), only the first 
$x$-moment of $Q_2^i$ is constrained. We simply assume that
$Q_2^i \propto Q_0^i$ and take
\begin{equation}
Q_2^i(x,0,\mu^2)=\beta^i \,Q_0^i(x,0,\mu^2) \,,
\label{eq:beta}
\end{equation}
where $\beta^i$ are constants. 
From Eq.~(\ref{eq:q02}), we obtain
\begin{equation}
\beta^i=\frac{6}{5} \frac{d^i(0)}{M_2^i(0)}+\frac{1}{2} \,,
\label{eq:beta2}
\end{equation}
which gives
\begin{equation}
\beta^u=-4.4\,, \quad \quad  \beta^d=-8.9\,, \quad \quad \beta^s=0.5 \,.
\label{eq:beta3}
\end{equation}
In this numerical estimate, we  assume that
$d^u=d^d \approx -2$ and $d^s \approx 0$, which agrees with the 
SU(3)-symmetric chiral quark soliton model calculation of the nucleon 
$D$-term~\cite{Kivel:2000fg}:
  $\sum_i d^i(0) \approx -4$,
and takes into account the SU(3) symmetry breaking in the nucleon PDFs
(the suppression of the strange quark PDF at the low resolution scale).
The momentum fractions $M_2^i$ were evaluated at $\mu_0=0.6$ GeV
 using the LO GRV parton PDFs~\cite{Gluck:1998xa}.

In general, $\beta^i$ also depend on $\mu^2$. 
However, 
as will be seen from the general expression for the DVCS amplitude,
at small values of $\xi$ typical for the HERA data on the total DVCS cross
section, the contribution of the $Q_2^i$ function is kinematically 
suppressed. Therefore, the goodness of the description of the data is not
affected by the exact values of $\beta^i$, and we simply used 
 Eq.~(\ref{eq:beta3})
 at all $\mu^2$.

Until recently, the $t$-dependence of the DVCS cross section was not 
measured. One would simply assume that the DVCS cross section exponentially
depends on $t$,
\begin{equation}
\frac{d \sigma_{{\rm DVCS}}(x_{Bj},Q^2,t)}{dt}=\exp\left(-B\,|t|\right)\left(\frac{d \sigma_{{\rm DVCS}}(x_{Bj},Q^2,t)}{dt}\right)_{t=0}
 \,,
\label{eq:expfit}
\end{equation}
such that the total DVCS cross section is
\begin{equation}
\sigma_{{\rm DVCS}}(x_{Bj},Q^2)=\frac{1}{B} \left(\frac{d \sigma_{{\rm DVCS}}(x_{Bj},Q^2,t)}{dt}\right)_{t=0}
 \,.
\label{eq:expfit2}
\end{equation}
The value of the slope parameter $B$ was rather uncertain,
 $5 \leq B \leq 9$ GeV$^{-2}$. The range of the values covers the
 experimentally measured range of the
$t$-slope of electroproduction of light vector mesons at HERA.
However, very recently the $t$-dependence of the total DVCS cross section
for $0.1 \leq |t| \leq 0.8$ GeV$^2$ and at $Q^2=8$ GeV$^2$ 
 was measured by the H1 collaboration
 at HERA and was fitted by the exponential form of Eq.~(\ref{eq:expfit})
with the result 
$B=6.02 \pm 0.35 \pm 0.39$ GeV$^{-2}$~\cite{Aktas:2005ty}.

In our numerical estimates of the DVCS cross section, we calculate 
the DVCS amplitude at $t=0$ and then use Eq.~(\ref{eq:expfit2}) in order 
to find the $t$-integrated DVCS cross section.
In general, the slope $B$ should decrease with increasing $Q^2$.
A particular model for the $Q^2$-dependent slope was 
suggested in~\cite{Freund:2002qf}: $B(Q^2)=8\,(1-0.15 \ln(Q^2/2))$ 
GeV$^{-2}$. In our analysis,  we use the same $Q^2$-dependence, 
\begin{equation}
B(Q^2)=7.6 \,\left(1-0.15 \ln(Q^2/2) \right) \quad {\rm GeV}^{-2} \,,
\label{eq:numslope}
\end{equation}
but with a slightly smaller constant $7.6$ GeV$^{-2}$, which 
 is chosen such that 
Eq.~(\ref{eq:numslope})  reproduces the H1 value of the slope 
at $Q^2=8$ GeV$^2$.

In summary, our parameterization of the GPDs $H^i$ is defined by
Eqs.~(\ref{eq:q02}), (\ref{eq:q0}) and (\ref{eq:beta}).  
The $t$-dependence of the DVCS cross section is given by Eqs.~(\ref{eq:expfit})
and (\ref{eq:numslope}).
This is the minimal version of the dual representation of the GPDs, which can
be readily extended by considering more $Q_k^i$ functions,
a more elaborate $t$-dependence and by taking into account the other GPDs of 
the proton. The main practical advantage of our representation is that
the $\mu^2$-evolution of $Q_{0,2}^i$ is given by the usual DGLAP
evolution of the singlet PDFs, see Eq.~(\ref{eq:q0}).

\section{Description of low-$x$ HERA data on DVCS cross section}
\label{sec:lowx}

In this section, we evaluate the total DVCS cross section using the minimal
model of the dual representation of the GPDs and compare the results to
the HERA data \cite{Chekanov:2003ya,Aktas:2005ty}.

The total unpolarized DVCS cross section on the photon level reads, see
e.g.~\cite{Freund:2001hm},  
\begin{equation}
\sigma_{{\rm DVCS}}(x_{Bj},Q^2)=\frac{\alpha_{e.m.}^2 x_{Bj}^2 
\pi \,(1-\xi^2)}{Q^4 \sqrt{1+4 x_{Bj}^2
m_N^2/Q^2}}\int_{t_{min}}^{t_{max}} dt \,|{\bar {\cal A}}_{{\rm DVCS}}(\xi,t,Q^2)|^2 \,,
\label{eq:cs1}
\end{equation}
where $\alpha_{e.m.}$ the fine-structure constant;  
$\xi =1/2 x_{Bj}/(1-x_{Bj}/2)$ in the Bjorken limit;
$t_{max} \approx 0$ and $t_{min} \approx -1$ GeV$^{-2}$;
$|{\bar {\cal A}}_{{\rm DVCS}}|^2$ is the squared and spin-averaged DVCS
amplitude.

To the leading order in $\alpha_s$, the DVCS amplitude is expressed
in terms of the singlet combination of the GPDs $H^i$,
\begin{equation}
{\cal A}_{{\rm DVCS}}(\xi,t,Q^2)=\sum_i e_i^2 \int^{1}_{0} dx \,
 H^i(x,\xi,t,Q^2)\left(\frac{1}{x-\xi+i0}+\frac{1}{x+\xi-i0}
\right) \,.
\label{eq:a}
\end{equation}
Using our model for the GPDs and the results of~\cite{Polyakov:2002wz},
 the DVCS amplitude can be
presented in a compact form in terms of the $Q_0^i$ and $Q_2^i$ functions
\begin{equation}
{\cal A}_{{\rm DVCS}}(\xi,t,Q^2)=-\sum_i e_i^2 \int^{1}_{0} \frac{dx}{x}  \sum_{k=0}^{2} x^k Q_k^i(x,t,Q^2) \left(\frac{1}{\sqrt{
1-\frac{2x}{\xi}+x^2}}+\frac{1}{\sqrt{1+\frac{2x}{\xi}+x^2}}-2
\delta_{k0} \right) \,.
\label{eq:a2}
\end{equation}

Using the exponential ansatz for the $t$-dependence of the DVCS cross section, 
the total DVCS cross section is expressed in terms of the
DVCS amplitude at $t=0$ [see Eq.~(\ref{eq:expfit2})]
\begin{equation}
\sigma_{{\rm DVCS}}(x_{Bj},Q^2)=\frac{\alpha_{e.m.}^2 x_{Bj}^2 
\pi \,(1-\xi^2)}{Q^4 \sqrt{1+4 x_{Bj}^2
m_N^2/Q^2}}\frac{1}{B(Q^2)}\,|{\bar {\cal A}}_{{\rm DVCS}}(\xi,t=0,Q^2)|^2 \,,
\label{eq:cs2}
\end{equation}
where ${\cal A}_{{\rm DVCS}}(\xi,t=0,Q^2)$ is given by Eq.~(\ref{eq:a2}) evaluated
 with $Q_{0,2}^i(x,0,Q^2)$.

\begin{figure}[t]
\begin{center}
\epsfig{file=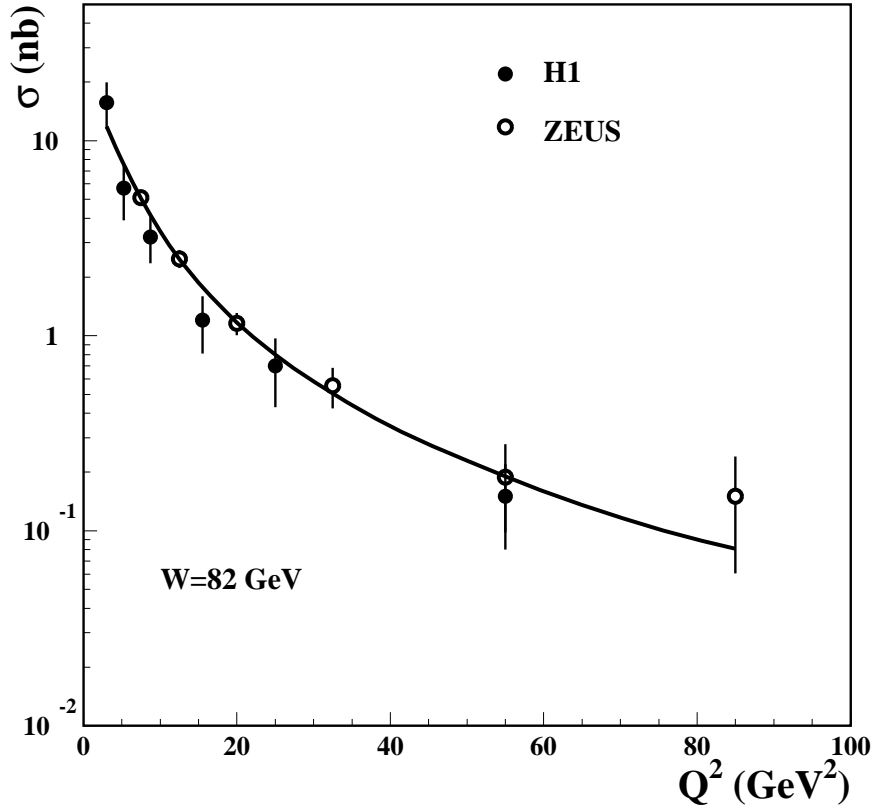,width=12cm,height=12cm}
\caption{The total DVCS cross section at $W=82$ GeV as a function of $Q^2$.
The predictions of the dual parameterization of the GPDs (solid curve)
is compared to the H1~\protect\cite{Aktas:2005ty} and 
ZEUS~\protect\cite{Chekanov:2003ya}. The error bars represent the statistical
 and systematic uncertainties added in quadrature.
}
\label{fig:q2}
\end{center}
\end{figure}

\begin{figure}[t]
\begin{center}
\epsfig{file=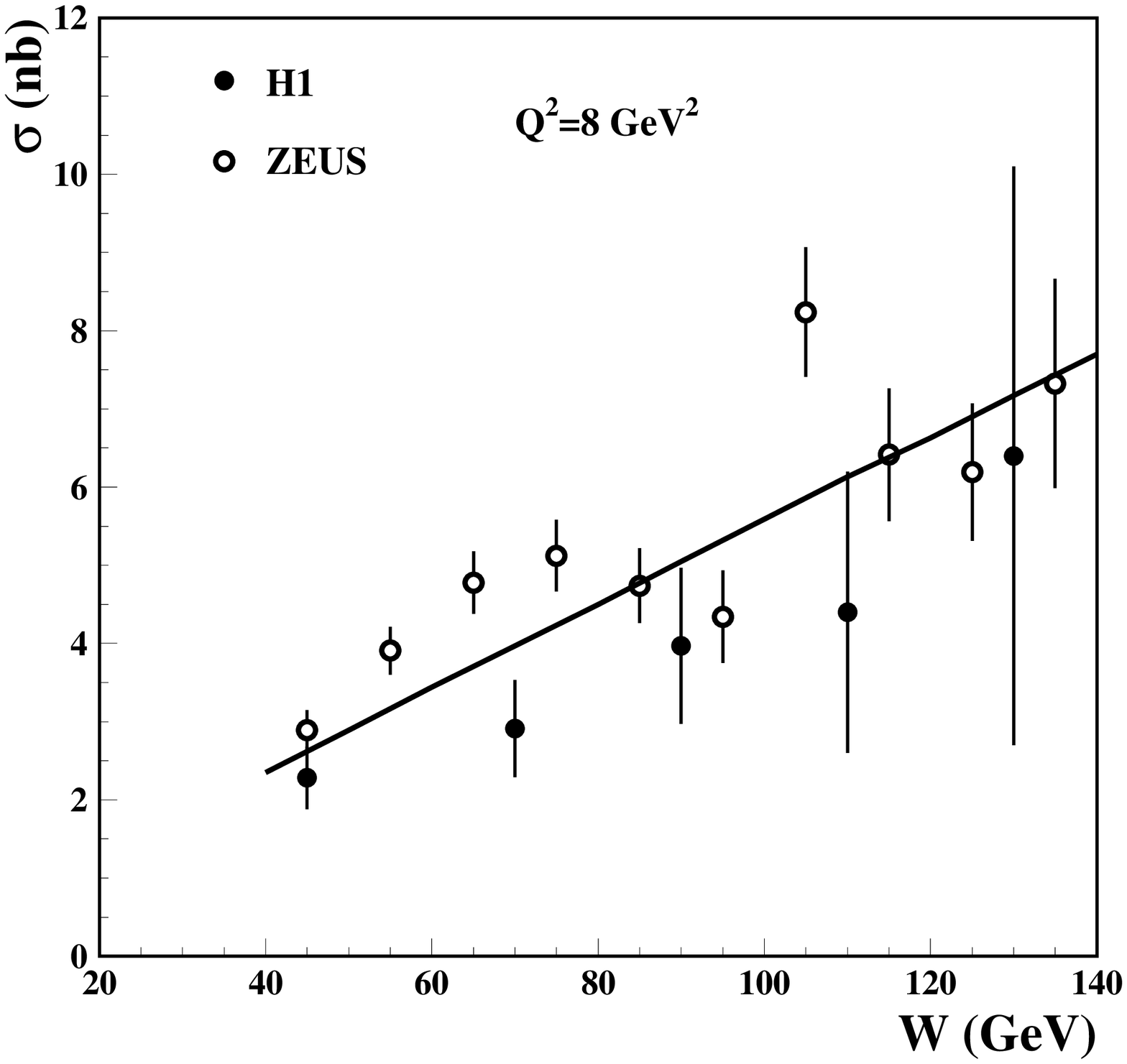,width=12cm,height=12cm}
\caption{The total DVCS cross section at $Q^2=8$ GeV as a function of $W$.
The predictions of the dual parameterization of the GPDs (solid curve)
is compared to the H1~\protect\cite{Aktas:2005ty} and 
ZEUS~\protect\cite{Chekanov:2003ya}. The error bars represent the statistical
 and systematic uncertainties added in quadrature.
}
\label{fig:w}
\end{center}
\end{figure}

Our predictions for the $Q^2$ and $W$-dependence of the
total DVCS cross section are presented in Figs.~\ref{fig:q2} and
\ref{fig:w}, respectively. For comparison, we also present
the H1~\cite{Aktas:2005ty} and ZEUS~\cite{Chekanov:2003ya} data.

Note that the ZEUS data points, which were taken at $W=89$ GeV and at 
$Q^2=9.6$ GeV$^2$, have been rescaled to the
H1 values of $W=82$ GeV and $Q^2=8$ GeV$^2$ using the fitted $W$ and 
$Q^2$-dependence of the DVCS cross section: $\sigma_{{\rm DVCS}} \propto W^{0.75}$ 
and $\sigma_{{\rm DVCS}} \propto 1/(Q^2)^{1.54}$~\cite{Chekanov:2003ya}.

For the proton forward PDFs, which are required to evaluate $Q_0^i$, we used
the LO CTEQ5L parameterization~\cite{Lai:1999wy}.

One can see from Fig.~\ref{fig:q2} that the absolute value and the
 $Q^2$-dependence of the total DVCS cross
section is described very well. The agreement with the data 
at the highest values of  $Q^2$ would have been worse, if we had used 
the $Q^2$-independent slope $B$.

From Fig.~\ref{fig:w} one can see that  the absolute value and 
the $W$-dependence of the 
DVCS cross section is also reproduced rather well. However, one should note
 a slight discrepancy between the ZEUS and H1 data points at lower values of
$W$ and large experimental errors at the high end of $W$. 

It is important to emphasize that our predictions for the total DVCS cross
section were made using the parameterization of the GPD, which contains
no free parameters (the role of $Q_2^i$ and $\beta$, see Eq.~(\ref{eq:beta}), is 
negligible in the H1 and ZEUS kinematics). It is very remarkable that
the agreement with the data is so good.

In order to understand, at least partially, the success of the dual
 parameterization of GPDs in the description of the low-$x$ HERA
data, it is instructive to analyze the DVCS amplitude ${\cal A}_{{\rm DVCS}}$
 of 
Eq.~(\ref{eq:a2}) in some detail.  Evaluating the imaginary and real parts of
Eq.~(\ref{eq:a2}), one obtains~\cite{Polyakov:2002wz} 
\begin{eqnarray}
&&{\rm Im}\, {\cal A}(\xi,Q^2)=-\sum_i e_i^2\int_a^1 \frac{dx}{x} \frac{1}{\sqrt{2x/\xi-x^2-1}} \sum_{k=0}^{2} x^k Q_k(x,0,Q^2) \,,\nonumber\\
&&{\rm Re}\, {\cal A}(\xi,t)=-\sum_i e_i^2\int_a^1 \frac{dx}{x} \sum_{k=0}^2 x^k Q_k(x,0,Q^2) \Big(\frac{1}{\sqrt{1+2x/\xi+x^2}} -2 \delta_{k 0}\Big)  \nonumber\\
&&-\sum_i e_i^2 \int^a_0 \frac{dx}{x} \sum_{k=0}^2 x^k Q_k(x,0,Q^2) \Big(\frac{1}{\sqrt{1-2x/\xi+x^2}}+\frac{1}{\sqrt{1+2x/\xi+x^2}} -2 \delta_{k 0}\Big) \,,
\label{eq:imre}
\end{eqnarray}
where $a=(1-\sqrt{1-\xi^2})/\xi$.

Al low $x_{Bj}$, $\xi \approx x_{Bj}/2$ and the integration limit is $a
\approx \xi/2 =x_{Bj}/4$.
Thus, the functions $Q_0^i$ and $Q_2^i$ are never sampled at $x < x_{Bj}/4$.
This is clearly an advantage over the double distribution parameterization
of GPDs, where the forward
parton distributions are sampled all the way down to $x=0$, which results
in the acute sensitivity to the
unmeasured, very low-$x$ behavior of the proton PDFs and leads to a
gross overestimate of the data~\cite{Freund:2001hd}.

In addition, Eqs.~(\ref{eq:imre}) are convenient for the numerical
 implementation
since the integrands do not contain large end-point contributions, as can be
explicitly seen by changing the integration variables.

\section{Discussion and Conclusions}
\label{sec:end}

We presented and discussed a new leading order parameterization of GPDs
introduced in~\cite{Polyakov:2002wz}. In its minimal form, the parameterization
is defined by the forward singlet quark PDFs and the form factors of
 the energy-momentum tensor,
 see Eqs.~(\ref{eq:q02}) and (\ref{eq:q0}). The $t$-dependence of
the DVCS cross section was assumed in a simple factorized form with 
the $Q^2$-dependent slope, see Eqs.~(\ref{eq:expfit}) and (\ref{eq:numslope}).

We showed that our parameterization of the GPDs describes very well the
absolute value,  the $Q^2$-dependence and $W$-dependence of the HERA data on
the total DVCS cross section. Moreover, since the data is at low $x_{Bj}$,
our parameterization can be simplified by omitting the contribution of the
$Q_2^i$ function. This means that we achieved a remarkably good description
of a large set of the data on DVCS using a parameterization of the GPDs
which contains no free parameters! 

We discuss that our parameterization suits the low-$x_{Bj}$ kinematics
 especially well
because the quark singlet PDFs are never probed at the unmeasurably low values
of Bjorken $x$, as is the case for the popular double distribution 
model~\cite{Freund:2001hd}, and because the expression for the DVCS amplitude
is numerically stable.
This allows us to advertise our model as a better alternative to the popular 
double distribution parameterization of the GPDs, at least in the low-$\xi$
region.

The parameterization presented in this work can be readily generalized
by including more $Q_k^i$ functions, considering the GPDs $E$, $\tilde{H}$ and
$\tilde{E}$ and by using more elaborate models of the $t$-dependence. This
was not necessary in the H1~\cite{Aktas:2005ty} and 
ZEUS~\cite{Chekanov:2003ya} kinematics, but
might be required for the HERMES and CLAS kinematics.
   
Also, the role of next-to-leading order (NLO) corrections and higher
twist effects should be investigated. In particular, it is important to
compare  the size of the NLO corrections using the dual parameterization
with the results of the  analysis using the double distribution
parameterization, where the NLO corrections were found to be 
large~\cite{Freund:2001hd}.

\section*{Acknowledgements}

We would like to thank M. Strikman and P. Pobylitsa for  valuable discussions
 and comments. It is a pleasure to thank A. Freund for carefully reading the
initial draft of the  manuscript and making valuable suggestions. This
work is supported by the Sofia Kovalevskaya Program of the Alexander
von Humboldt Foundation.

\end{document}